\documentclass[article,superscriptaddress,pt11,twocolumn]{revtex4-1}
\pdfoutput=1
\usepackage{graphicx}
\usepackage{dcolumn}
\usepackage{amsmath}
\usepackage{amsfonts}
\usepackage{upgreek}
\usepackage{color}
\usepackage[normalem]{ulem}
 \usepackage{caption}

\begin{document}

\title{\textsf{Emergent electromagnetism induced by topological defects created during magnetization reversal in nanowires}}

\author{M. Charilaou}
\email[Email: charilaou@mat.ethz.ch]{}
\affiliation{Laboratory of Metal Physics and Technology, Department of Materials, ETH Zurich, Zurich 8093, Switzerland}
\author{H.-B. Braun}
\email[Email: beni.braun@ucd.ie]{}
\affiliation{School of Physics, University College Dublin, Dublin 4, Ireland}
\author{J. F. L\"offler}
\affiliation{Laboratory of Metal Physics and Technology, Department of Materials, ETH Zurich, Zurich 8093, Switzerland}

\begin{abstract}
We report that the irreversible magnetization switching process in ferromagnetic nanoparticles is governed by the formation and dynamics of topological point-defects in the form of hedgehog-antihedgehog pairs. After nucleation, these pairs rapidly separate with speeds exceeding domain wall velocities, and they generate an emergent electric field of solenoidal character and substantial magnitude. 
\end{abstract}
\maketitle

\section{Introduction}

Topological concepts allow for the identification of extraordinarily stable magnetization textures that play the role of quasiparticles which can be  created and controlled individually \cite{milde2013,sampaio2013,hoffmann2017}. Smooth topological defects such as domain-wall pairs in easy-plane or skyrmions in easy-axis ferromagnets are characterized by the fact that upon stereographic projection the magnetization field wraps once around the circle or sphere, respectively. They are hence endowed with topological stability as they cannot be continuously deformed into the uniform state and as a consequence they are considered promising candidates for information carriers in prospective racetrack-type memories \cite{parkin2008,sampaio2013}. They may, however, decay via the formation of point-like singularities, 
such as hedgehogs or Bloch points or monopoles \cite{milde2013,schuette2014,nagaosa2016,charilaou2017}. Thus the formation and dynamics
of topological point defects must play a central role in magnetization processes.

The judicious ability to control magnetization reversal in nanostructures underpins the progress of magnetic technologies \cite{parkin2015,sander2015}. Magnetic switching may become surprisingly rich as dimensions become comparable to fundamental magnetic length-scales even for nanoparticles that are in a single-domain state at remanence. Conventional wisdom associates magnetization reversal in cylindrical nanoparticles with symmetric curling-type processes \cite{aharoni2000,pacheco2017}, but these considerations neglect the global topological constraints which preclude complete reversal via continuous processes. 

In this article we show that irreversibility of the magnetization process is directly linked to the pair-creation of topological point defects in the form of hedgehog-antihedgehog pairs, whose subsequent separation accomplishes the complete reversal of the magnetization. These fast moving hedgehogs produce a solenoidal emergent electric field, in precise analogy to the magnetic field of a moving electric charge, and it has a strength of the order of MV/m. Hence, the formation and dynamics of topological point defects plays a crucial role in magnetization processes in some of the most extensively studied nano-scale objects, namely magnetic nanoparticles that are in a single-domain state at remanence \cite{krishnan2016,guimaraes2017}, which are highly relevant for spintronics \cite{bruno2004,pacheco2017,hoffmann2017}, data-storage \cite{sun2000}, and biomedicine \cite{krishnan2016,myrovali2016}. 

Specifically, using high-resolution micromagnetic simulations (cf. Methods) we find that in elongated nanoparticles of simple ferromagnets, such as permalloy or cobalt, switching processes initiated via a curling instability occur in two stages. In a first stage this instability develops into the formation of a skyrmion line. This is fully reversible and hence upon removal of the applied field, the magnetization returns to its initial state. The second stage is irreversible and involves the breaking of the skyrmion line with concomitant creation of topological point defects in the form of hedgehog-antihedgehog pairs, which after creation rapidly separate with velocities that significantly exceed that of domain walls \cite{parkin2015b}. Being topologically nontrivial objects \cite{braun2012}, these rapidly moving hedgehogs give rise to significant forces on the conduction electrons which are described by emergent electric fields.

\begin{figure*}
	\centering
		\includegraphics[width=1.8\columnwidth]{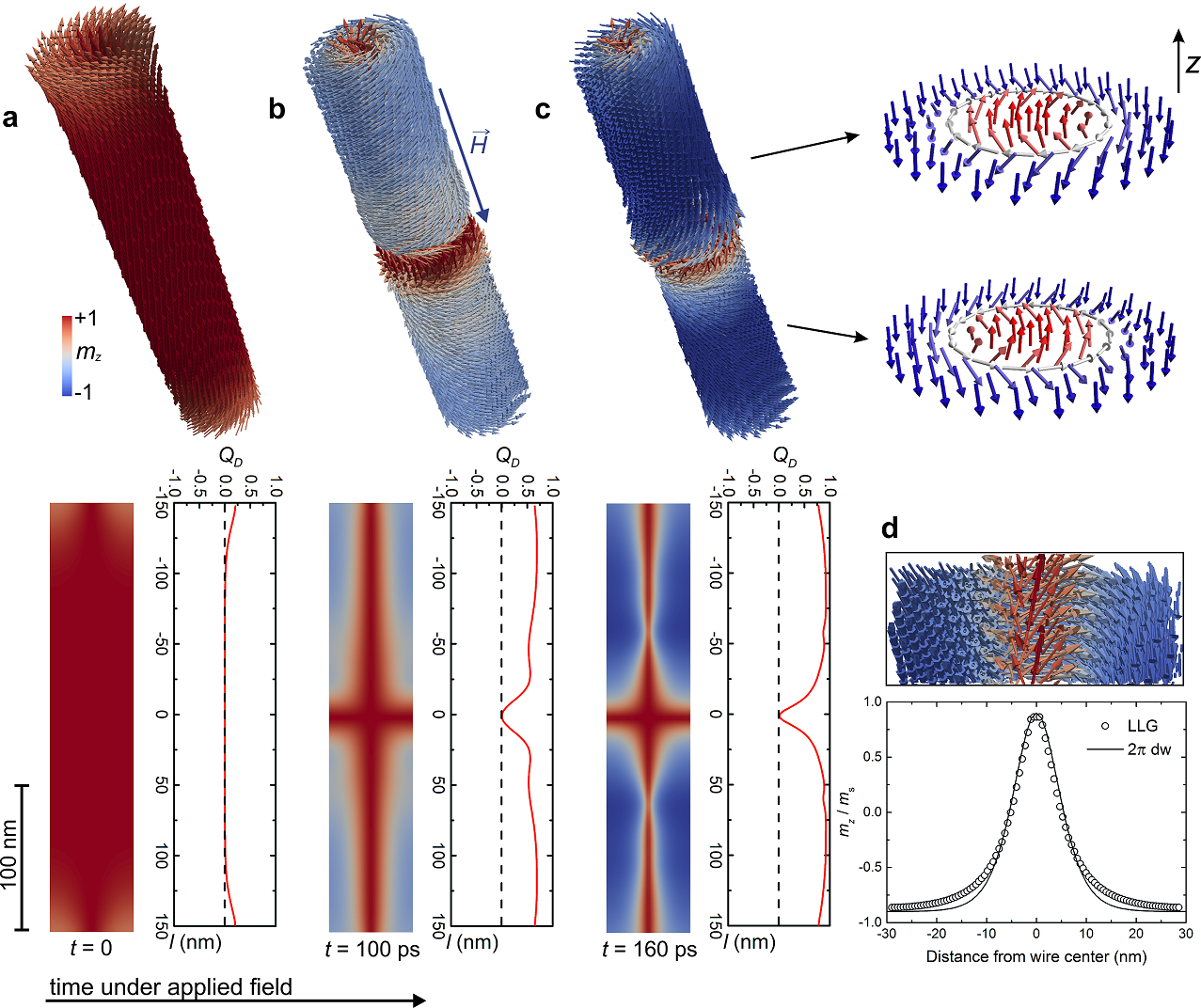}
	\caption{
	\textbf{Formation of skyrmion lines during the magnetization process.} {\bf a} - {\bf c}, Snapshots of a micromagnetic simulation showing the initial, reversible, stages of the switching process in a cylindrical nanowire of finite length: {\bf a}, at $0s$, {\bf b}, at $100 \, {\rm ps}$ and, {\bf c}, at  $160 \,{\rm ps}$, 
after having applied a constant field of $125~{\rm mT}$ along $-z$. The top panels show vector-field plots and the bottom panels show contour plots of the $z$-component of the magnetization together with the topological charge density integrated across the cylinder. The initial curling-mode instability develops into the formation of two skyrmion lines of opposite handedness. {\bf d}, The profile across the wire is well described by a variational ansatz \cite{braun1994,romming2015}.
	}
	\label{fig1}
\end{figure*}

\section{Methods}

The total energy density of the system consists of contributions from ferromagnetic exchange, uniaxial anisotropy, Zeeman energy, and dipolar interactions

\begin{equation}
F= A \left(\nabla \mathbf{m}\right)^2-K_\mathrm{u}  m_z^2-\mu_0 M_\mathrm{s}\mathbf{H}_\mathrm{ext}\cdot \mathbf{m}-\frac{\mu_0 M_\mathrm{s}}{2}\mathbf{m} \cdot \mathbf{h}_\mathrm{d},
\end{equation}

\noindent where $A$ is the exchange stiffness, $\mathbf{m}$ is the magnetization unit vector ($\mathbf{m}=\mathbf{M}/M_\mathrm{s}$ with $M_\mathrm{s}$ the saturation magnetization), $K_\mathrm{u}$ is the anisotropy energy density, $\mathbf{H}_\mathrm{ext}$ is the external field, and $\mathbf{h}_\mathrm{d}$ denotes the local demagnetizing field. 

We solve the Landau-Lifshitz-Gilbert equation of motion $\partial_t  \mathbf{m}= -\upmu_0\gamma \left(\mathbf{m} \times \mathbf{h}_\mathrm{eff}\right)+\alpha \left(\mathbf{m} \times \partial_t \mathbf{m}\right)$, where $\alpha$ is the dimensionless damping parameter, $\gamma$ is the electron gyromagnetic ratio, and $\mathbf{h}_\mathrm{eff}=-\partial_\mathbf{m} F/\upmu_0 M_\mathrm{s}$ is the effective magnetic field consisting of both internal and external fields.

For the calculations the graphics-processing-unit accelerated software package MuMax3 \cite{mumax3} was used in the high-damping case with $\alpha=0.1$. Occasional checks with $\alpha=0.01$ and 0.001 were made to test the effects of damping on the simulation findings, but no significant change in the results was found.  

%The variational ansatz for a 2$\pi$ domain wall as discussed in Ref. \cite{braun2012} is $\cot(\theta/2)=\cosh(R)/\sinh(\rho/\delta)$, where $\theta$ is the angle of the magnetic moment at a distance $\rho$ from the wire center, $R$ is a variational parameter, and  $\delta_\mathrm{s}$ is the skyrmion radius.

%$\delta_m\approx 8$~nm

The simulations shown here were for permalloy (Fe$_{20}$Ni$_{80}$) with $A=13$~pJ/m and $M_\mathrm{s}=800$~kA/m. We chose this material because it is commonly synthesized in laboratories in form of nanoparticles and nanowires, which makes the experimental realization highly feasible. In addition we considered a wide range of materials, with different values of the uniaxial anisotropy $K_\mathrm{u}$, and found that the phenomena presented are valid when the diameter of the cylinder is comparable to the domain-wall width $\pi\delta_\mathrm{dw}$ (data not shown here).

The simulated nanowires shown in this paper have a diameter of 60 nm and length of 300 nm. For those simulations the cell size is set to 1.87 nm x 1.87 nm x 2.35 nm. Smaller and larger cell sizes were also implemented to verify the numerical stability of the results, with the cell size never being larger than 0.5$\delta_m=\sqrt{A/\mu_0 M_\mathrm{s}^2}\approx 4$ nm or smaller than 1 nm. Even though small cell sizes were used, quantum-mechanical effects were not considered in our simulations. 

\section{Results and discussion}

The nanowires that we study here are beyond the Stoner-Wohlfarth regime, where all particle dimensions are below the intrinsic length scales of the nanoparticle, i.e., the exchange length and the domain wall width, and the magnetization remains uniform even during switching \cite{wernsdorfer2001}. Here we consider elongated particles with diameter comparable to $2\pi \delta_m$.

In a cylindrical nanowire of permalloy, with a high aspect ratio (in this example 5:1), the equilibrium magnetization nearly equals the saturation value, only slightly reduced due to dipolar-induced curling at the ends of the cylinder. Once we apply a constant external field opposing the remanent state, the two ends act as nucleation sites. With increasing field strength and duration, curling becomes more pronounced when two propagation fronts with opposite chirality are generated at each end of the cylinder in order to reduce the dipolar energy of the system. Figure \ref{fig1} shows snapshots of micromagnetic simulations of a cylindrical nanowire with diameter 60 nm and length 300 nm in the presence of a constant external magnetic field. As time progresses spins on the outer rim of the wire align with the external field, whereas spins in the center point along the original remanent direction. Importantly, the spin structure in the middle section remains collinear, with all spins pointing in the original direction, i.e., opposing the external field. 
\begin{figure}
	\centering
		\includegraphics[width=1.0\columnwidth]{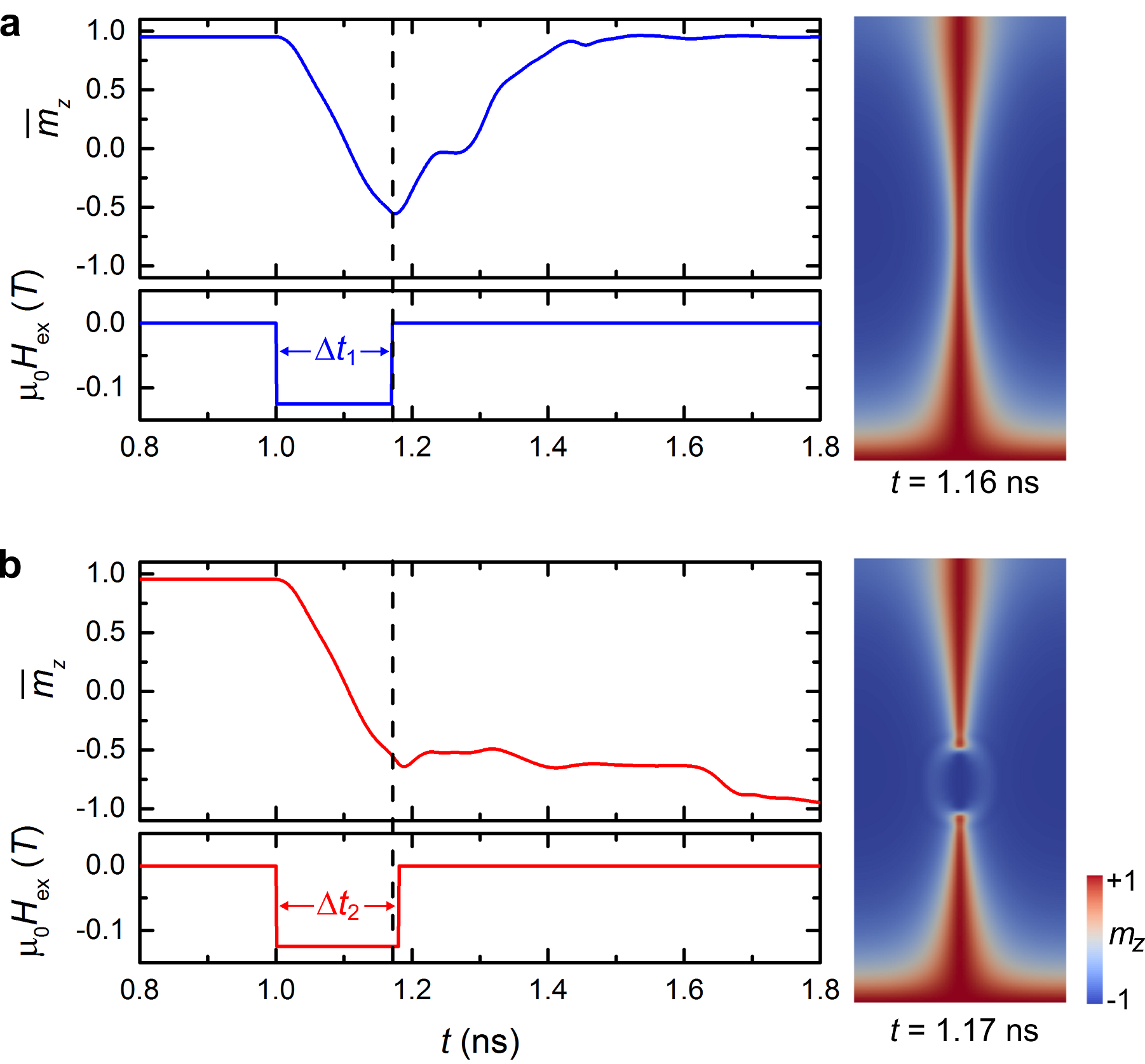}
	\caption{
	\textbf{Pair creation of topological point defects and the onset of irreversibility.} Time evolution of the average magnetization $\bar m_z$ during a field pulse, and the contour plot of $m_z$ at the time when the applied magnetic field is switched-off. {\bf a}, The magnetization process is reversible if the applied field is switched-off before the skyrmion lines break ($\Delta t_1 = 0.16$~ns). {\bf b}, Irreversibility sets in only when the skyrmion line is destroyed via the creation of  a hedgehog-antihedgehog pair before switching-off the applied field ($\Delta t_2 = 0.17$~ns). Only in the latter case full magnetization reversal occurs ($\bar m_z \approx -1$). 
}
	\label{fig2}
\end{figure}
The magnetization profile of this state fits that of a 2$\pi$-domain wall profile \cite{braun1994,braun2012}, which in fact is an excellent approximation for the profile of a \textit{skyrmion}: $\cot(\theta/2)=\cosh(R)/\sinh(\rho/\delta)$, where $\theta$ denotes the polar angle of the local magnetic moment, $\rho$ the distance from the skyrmion center, $\delta$ the intrinsic length scale, and $R$ the dimensionless skyrmion radius. Together with the azimuthal angle $\phi$, the polar angle $\theta$ parametrizes the magnetization unit vector via ${\bf m} = ( \sin \theta \cos \phi, \sin \theta \sin \phi, \cos \theta)$. If the azimuthal angle $\phi$ describes a N\'eel or a Bloch wall configuration (cf. Fig. \ref{fig1}c), this texture has a non-trivial topology reflected by a non-zero topological charge \cite{braun2012}, defined via 

\begin{align}
4 \pi Q_M &= \int_M \!\!\! \epsilon_{abc} m^a \, dm^b \wedge dm^c =  \int \! du \; dv \,\sin \theta \, J(\theta,\phi)
\label{eq:winding}
\end{align}
where $J(\theta,\phi) = \partial_u \theta \partial_v \phi -  \partial_v \theta \partial_u \phi$. In the following we shall consider two different situations where the real-space manifold $M$ parametrized by $u,v$ is either a disk $D$ across the wire, or a sphere $S^2$ around any point in the interior of the sample. In the wires shown in Fig. \ref{fig1}c, $Q_D \simeq 1$, except at the regions close to the top and bottom due to the twisting of the moments on the surface. This state corresponds to an extended  \emph{skyrmion line} \cite{braun1999} which broadens to a nearly uniform state near the middle of the sample. This middle uniform segment shrinks with increasing field duration and strength.
 
\begin{figure*}[t]
	\centering
		\includegraphics[width=2.0\columnwidth]{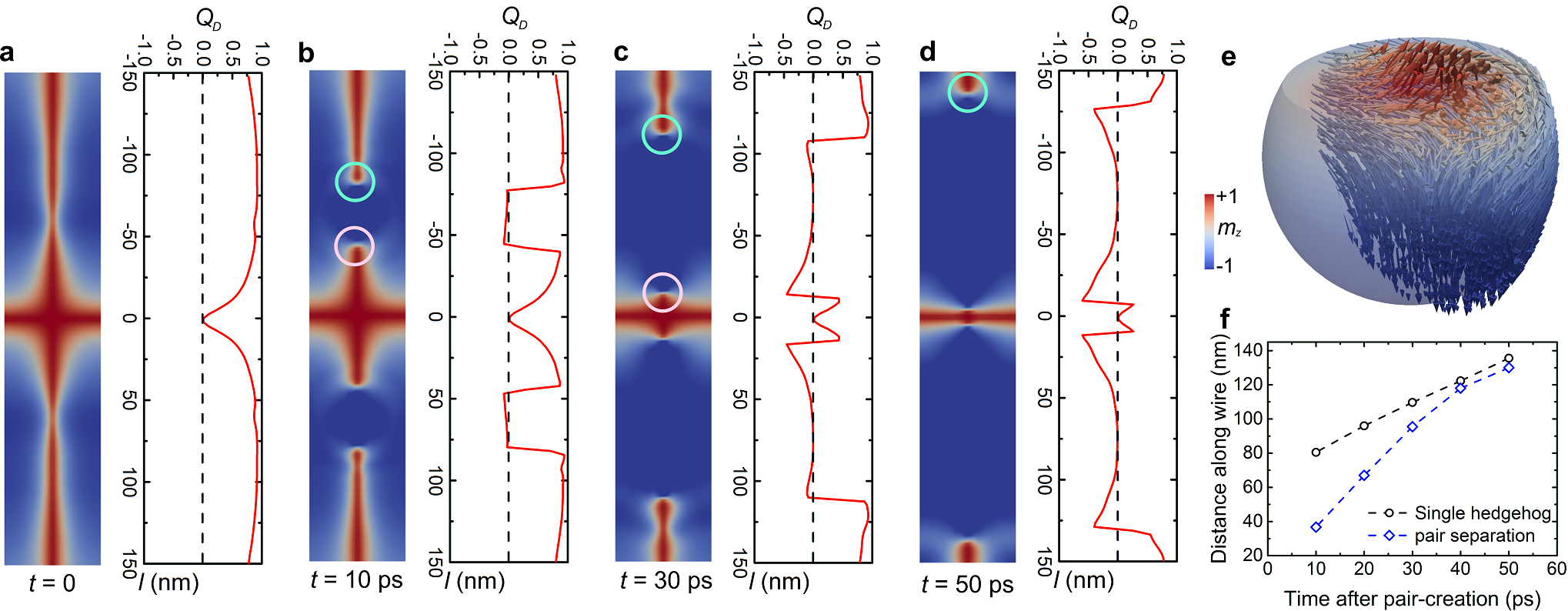}
	\caption{\textbf{Dynamics of separating hedgehog-antihedgehog pairs.} Contour plots of $m_z$ and $Q_D$ as a function of $z$ at, {\bf a}, 
	$t=0$, {\bf b}, $t=10~{\rm ps}$, {\bf c}, $t=30~{\rm ps}$, {\bf d}, $t=50~{\rm ps}$ after the field has been switched-off. The circles in the contour plots indicate the position of the $Q_{S^2}=1$ (cyan) hedgehog and the $Q_{S^2}=-1$ (pink) antihedgehog as a function of time. {\bf e}, Detailed view of the moment distribution of a hedgehog at the end of the skyrmion line. {\bf f}, Hedgehog separation (circles) and location of the top hedgehog (diamonds) as a function of time yielding a maximal relative velocity of $\approx 3\times 10^3 {\rm m/s}$.
	}
	\label{fig3}
\end{figure*}

At the critical value of the external magnetic field the dynamics exhibits critical slowing down, and switching will occur only after infinitely long field exposure. For super-critical fields the switching process depends crucially on the duration of the applied field. As shown in Fig. \ref{fig2}, if the field is switched off before the skyrmion lines break, the process is fully reversible and no magnetization switching occurs. In fact, the magnetization returns to its original remanent state within a fraction of a nanosecond, reminiscent of a radial exchange spring. Importantly, the state shown in Fig. \ref{fig2}a does not contain any point defects, i.e., $Q_{S^2}=0$ everywhere. Irreversible magnetization switching only occurs if the field is applied long enough to break the skyrmion lines, which entails the pair-creation of point defects. Such pair-creation abruptly introduces singular topological defects in the form of hedgehogs, or Bloch points, as shown in Fig. \ref{fig3}e, which are characterized by $Q_{S^2}=\pm 1$. Hence, we find that the topological charge $Q_{S^2}=0$ ($\neq 0$) is a direct measure of the reversibility (irreversibility) of the switching process.

This change of topology generates a new dynamic magnetic state in the nanowire. The hedgehogs rapidly move away from each other, as seen in Fig. \ref{fig3}b--d which shows their separation as a function of time. In fact, the relative velocity of the hedgehogs is around 2900 m/s, which is of the order of phonon velocities. 

\begin{figure*}
	\centering
		\includegraphics[width=1.5\columnwidth]{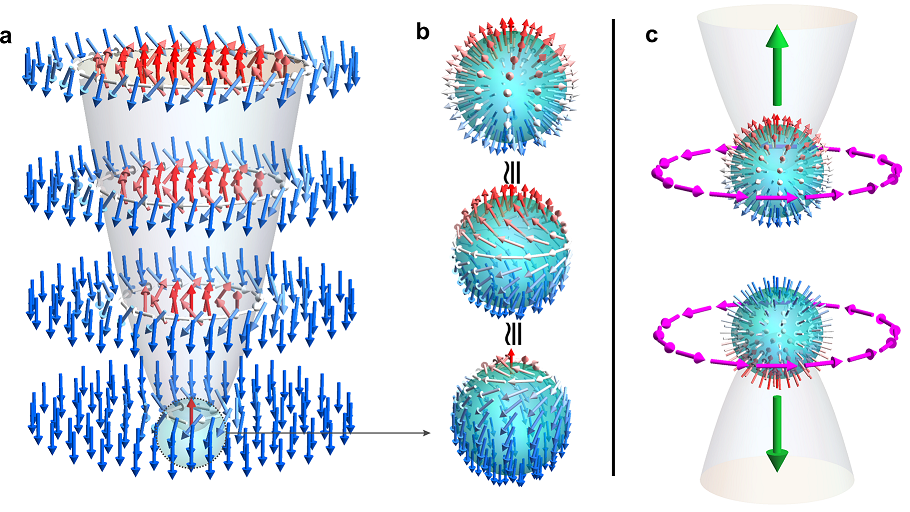}
	\caption{\textbf{Emergent electric field of moving hedgehogs or monopoles.} {\bf a}, Schematic view of the end of a skyrmion line. {\bf b}, The corresponding magnetization texture is seen to be topologically equivalent to a hedgehog with radially outward oriented magnetization ($Q_{S^2}=1$). {\bf c}, A pair of separating hedgehogs of opposite topological charge ($Q_{S^2}=\pm 1$) gives rise to an emergent solenoidal electric field (purple), in analogy to the magnetic field created by a pair of separating electric charges of opposite sign. For the process shown in Fig. \ref{fig3}, this emergent electric field of such a  magnetic monopole is of the order of  0.45 MV/m.}
	\label{fig4}
\end{figure*}

Importantly, the movement of the hedgehogs along the wire produces an \emph{emergent electric field}, as illustrated in Fig. \ref{fig4}. It is known that moving topologically non-trivial spin textures may give rise to weak emergent electric fields, which act on the conduction electrons. This effect, however, has only been considered for materials exhibiting Dzyaloshinskii-Moriya interactions \cite{milde2013,schuette2014,nagaosa2016,lin2016}, and the strength and character of the emergent field have remained unknown. In our simulations we consider materials that are simple ferromagnets, and the formation of the skyrmion lines and consequently of the hedgehogs is a dynamic effect produced solely from the topology of the magnetic spin texture in the cylindrical nanowires. Further, the ultrafast movement of the hedgehogs in a straight line has to our knowledge never been observed, and it has striking consequences: the emergent electric field due to the hedgehog's linear movement has solenoidal character, in analogy to the magnetic field of a moving electric charge.

The components of the emergent electric field are given by \cite{volovik1987,rosch2012}

\begin{equation}\label{efield}
E_i^{\rm em}  = \hbar \, {\bf m} \cdot (\partial_i {\bf m} \times \partial_t {\bf m}) \; .
\end{equation}
For electrons aligned with magnetization parallel to the magnetization unit vector ${\bf m}$ (i.e. spin antiparallel to {\bf m}), the emergent electric charge is given by \cite{rosch2012} $q^{\rm em}_\downarrow = -1/2$ and for the opposite spin $q^{\rm em}_\uparrow =+1/2$. Thus the net force on a conduction electron due to the emergent $E$-field is given by ${\bf F} = q^{\rm em} {\bf E}^{\rm em}$. For the spin texture of a vertically moving hedgehog, as shown in Fig. \ref{fig3}, the emergent field (\ref{efield}) has a solenoidal character, as shown in Fig. \ref{fig4}, and the magnitude of the force is then given by $F \approx \hbar v/2\lambda^2$. Here $\lambda$ is the characteristic length of the Bloch point spin texture and $v$ its velocity. For the values obtained from our simulations ($v=1360$~m/s and $\lambda \approx \delta \simeq 1~{\rm nm}$), we obtain $F \approx \hbar v/2\lambda^2  v  \approx  7.2 \times 10^{-14} {\rm J/m}$, which corresponds to a force exerted by a real electric field with magnitude $E_{\rm real} \approx   0.45 {\rm MV/m}$. This value is strikingly large, less than one order of magnitude smaller than the dielectric breakdown in vacuum ( $\approx 3 \; {\rm MV}/{\rm m}$), and it is produced by a single moving hedgehog. This suggests that moving or oscillating hedgehogs will emit measurable radiation due to the accelerating electrons and we propose that this could be detected in pump-probe switching experiments.

\section{Conclusions}

In conclusion, we have found that the magnetization switching in ferromagnetic nanoparticles is directly linked to the formation and dynamics of topological point-defects. In the example shown here, the switching occurs via the formation of two skyrmion lines. As long as the skyrmion lines are intact, the switching is fully reversible upon switching off the external magnetic field. As soon as the skyrmion lines break, however, hedgehog-antihedgehog pairs are created and the switching becomes irreversible. The rapidly moving hedgehogs generate an emergent electric field with substantially high magnitude and a solenoidal character. Hence, non-trivial topological spin textures and magnetic monopoles can exist dynamically even in simple systems, and their motion produces emergent electromagnetic fields.

\begin{acknowledgments}
MC and JFL gratefully acknowledge funding from ETH Zurich and the Swiss National Science Foundation (Grant No. 200021--172934) and HBB thanks the Science Foundation Ireland (Grant No. 11-PI-1048).
\end{acknowledgments}

\end{document}